\documentclass{PoS}
\usepackage{amsfonts,amssymb,graphicx}
\usepackage[tight]{subfigure}
\usepackage{rotating}
\def\epsfig{\psfig}
\usepackage{longtable}

\newcommand{\sgn}{{\rm sgn}}

\def\Z{\mathbb{Z}}

\def\ov{\overline}
\def\N{\mathbf{N}}
\def\Sym{\mathbf{Sym}}
\def\Anti{\mathbf{Anti}}
\def\Adj{\mathbf{Adj}}
\def\R{\mathbf{R}}

\def\ov{\overline}
\def\1{{\bf 1}}
\def\2{{\bf 2}}
\def\3{{\bf 3}}
\def\4{{\bf 4}}
\def\6{{\bf 6}}
\def\OR{\Omega\mathcal{R}}

\def\pp{\uparrow\uparrow}

\title{Towards exact field theory results for the Standard Model on fractional D6-branes}

\ShortTitle{Towards exact field theory results for the Standard Model on fractional D6-branes}

\author{\speaker
{Gabriele Honecker}\thanks{ The work of G.H. is partially supported by the 
{\it Research Center Elementary Forces and Mathematical Foundations (EMG)} at JGU Mainz.}\\
        Institut f\"ur Physik (WA THEP)\\ Johannes-Gutenberg-Universit\"at\\ D- 55099 Mainz, Germany\\
        E-mail: \email{Gabriele.Honecker@uni-mainz.de}}

\abstract{Fractional D6-branes on toroidal orbifold backgrounds are known to be able to accommodate the particle spectrum and gauge group of the Standard 
Model, but up to now exact results for their low-energy effective action are missing. Here we discuss how the conceptual ansatz of matching the string
theoretic gauge couplings at one-loop  with the supergravity expressions is generalised  from the six-torus to orbifold backgrounds on which the 
Standard Model spectrum can be realised on fractional D6-branes. The K\"ahler metrics and perturbatively exact holomorphic gauge kinetic functions 
can be classified in terms of the vanishing of some intersection angle and the related beta function coefficients, which potentially opens the possibility 
to extrapolate to smooth Calabi-Yau
backgrounds.
}

\FullConference{The XXIst Europhysics Conference on High Energy Physics-HEP 2011\\
		July 21-27 2011\\
		Grenoble, Rh\^{o}ne-Alpes, France
\hspace{75mm}
{\rm MZ-TH/11-31}}

\begin{document}
\section{Introduction}

Fractional D6-branes on $T^6/\Z_{2N}$ orientifold backgrounds of type IIA string theory have shown to be able to accommodate the
Standard Model (SM) spectrum~\cite{Gmeiner:2007zz,Honecker:2004kb}, and improved models are expected on rigid 
D6-branes on $T^6/\Z_2 \times \Z_{2M}$ with discrete torsion~\cite{Blumenhagen:2005tn}. While extensive searches for SM
spectra have been performed on a variety of toroidal orbifold backgrounds, the derivation of exact field theoretic results 
has to date focussed on the most simple case, the six-torus. We present here first results on the perturbatively exact holomorphic
gauge kinetic functions and on the leading order K\"ahler metrics on type IIA  $T^6/\Z_{2N}$ orientifold backgrounds by
matching conformal field theory (CFT) results on gauge thresholds at one string-loop with standard supergravity expressions.

\section{K\"ahler metrics and holomorphic gauge kinetic functions at one-loop}

 To study the low-energy effective field theory, the gauge couplings at one-loop can be computed using the magnetic background field method
(see~\cite{Lust:2003ky} for intersecting D6-branes on the six-torus and~\cite{Gmeiner:2009fb} on $T^6/\Z_{2N}$),
\begin{eqnarray}\nonumber
&\left(\frac{M_{\rm Planck}}{M_{\rm string}}\right)^2 {\rm Vol}({\rm D6}_a) &
+ \int_0^{\infty} dl \; l^{\varepsilon} \frac{\partial^2}{\partial {B_{\rm mag}}^2} \Bigl[ \langle {\rm D6}_a(B_{\rm mag}) |
 e^{- l \pi  H_{\rm closed}}| \Bigl(\sum_b |{\rm D6}_b\rangle + |{\rm O6}\rangle\Bigr)\Bigr]_{B_{\rm mag}=0}
\\
&\stackrel{\frac{1}{\varepsilon}=\ln \left(\frac{M_{\rm string}}{\mu}\right)^2}{=}&\hspace{-8mm}
\frac{1}{g_{a,{\rm tree}}^2}
+\frac{b_a}{16\pi^2} \ln\left(\frac{M_{\rm string}}{\mu}\right)^2 + \frac{\Delta_a}{16\pi^2} =\frac{1}{g_a^2(\mu)}
\quad {\rm with} \;\;
\begin{array}{c} b_a = \sum_b b_{ab}^{\mathcal{A}} + b_{aa'}^{\mathcal{M}},\\
\Delta_a=\sum_b \Delta_{ab}^{\mathcal{A}} + \Delta_{a,\Omega\mathcal{R}}^{\mathcal{M}}
,
\label{Eq:g_1-loop-string}
\end{array}
\end{eqnarray}
with the beta function coefficients $b_a$ of the gauge group $G_a$
and gauge thresholds $\Delta_a$ due to massive strings decomposed into 
contributions from individual open string sectors with annulus ($\mathcal{A}$) and M\"obius strip ($\mathcal{M}$) topology
and endpoints on D6-branes $a$ and $b$, the latter also including orbifold and orientifold images $(\theta^m a)_{m=0\ldots N-1}$ and $(\theta^n a')_{n=0\ldots N-1}$
on $T^6/\Z_{2N}$.

\noindent{\bf Matching  1-loop in string theory and supergravity.}
To obtain the canonical supergravity formulation, the string theoretic expression~(\ref{Eq:g_1-loop-string}) needs to be matched 
with the field theoretic one,
\begin{eqnarray}
\frac{1}{g_a^2(\mu)}&=& \Re({\rm f}_a) +\frac{b_a}{16\pi^2} \left[ \ln\left(\frac{M_{\rm Planck}}{\mu}  \right)^2 \!\!\!+  {\cal K} \right]
+ \frac{C_2(\Adj_a)}{8\pi^2} \; \left[{\cal K} \!- \ln g_a^{2}(\mu^2) \right]
- \sum_a \frac{C_2({\bf R}_a)}{8\pi^2} \; \ln  K_{{\bf R}_a}(\mu^2)
,
\nonumber
\end{eqnarray}
containing the holomorphic gauge kinetic function ${\rm f}_a$, the K\"ahler potential $\mathcal{K}$ and K\"ahler metrics $K_{{\bf R}_a}$
of the matter representations ${\bf R}_a \in \{(\N_a,\ov{\N}_b),(\N_a,\N_b), \Anti_a,\Sym_a,\Adj_a\}$ of $U(N_a) \times U(N_b)$
with quadratic Casimirs $C_2({\bf R}_a)$.
Using an iterative procedure (see~\cite{Akerblom:2007uc} for $h_{21}^{\rm bulk}=3$ on the six-torus and partial results on $T^6/\Z_2 \times \Z_2$ 
with discrete torsion), 
the K\"ahler potential for the field theoretical dilaton $S$ and bulk complex structure moduli $U_l$ and K\"ahler moduli $v_i$ on $T^6/\Z_{2N}$ and
$T^6/\Z_2 \times \Z_{2M}$ without and with discrete torsion
takes the form ~\cite{Honecker:2011sm},
\begin{eqnarray}
\mathcal{K}_{{\rm bulk}} = - \alpha \ln S - \alpha \sum_{l=1}^{h_{21}^{\rm bulk}} \ln U_l - \sum_{i=1}^3 \ln v_i
\qquad {\rm with} \quad
\alpha=1,2,4, \quad {\rm for} \quad h_{21}^{\rm bulk}=3,1,0
,
\nonumber
\end{eqnarray}
for various numbers $h_{21}^{\rm bulk}$ of bulk complex structures $U_l$.
At tree level, the holomorphic gauge kinetic function can be brought to the form
${\rm f}_a^{\rm tree}= S \,\tilde{X}_a^0 - \sum_{i=1}^{h_{21}^{\rm bulk}} U_i \, \tilde{X}_a^i$,
where $\tilde{X}_a^i$ are the (suitably normalised) bulk wrapping numbers of orientifold even three-cycles, 
$\Pi_a= \sum_{i=0}^{h_{21}} \left(\tilde{X}_a^i \Pi_i^{\rm even} + \tilde{Y}_a^i \Pi_i^{\rm odd}\right)$, specified in~\cite{Honecker:2011sm} for all orbifold 
backgrounds.
The term proportional to $C_2(\Adj_a)$ in the supergravity expression
contributes only to the matching of the $aa$ sector, i.e. strings with both endpoints on the same stack of D6-branes.
Analogously to the beta function coefficients and gauge thresholds, the one-loop contributions to the holomorphic gauge kinetic function form a sum 
over contributions from different open string sectors, \mbox{$\delta_{\rm total}^{\rm loop} {\rm f}_a = \sum_b \delta_b^{{\rm loop},{\cal A}} {\rm f}_a + 
\delta_{a'}^{{\rm loop},{\cal M}} {\rm f}_a$}.
The open string K\"ahler metrics take an universal shape for all factorisable six-torus and toroidal orbifold backgrounds~\cite{Honecker:2011sm} as displayed in 
table~\ref{Tab:Kaehlermetrics+delta_f_Annulus}, which fits with the alternative derivation on the
six-torus by means of disc scattering amplitudes~\cite{Lust:2004cx}.
\begin{table}
\begin{tabular}{|c||c|c|}\hline
$(\phi_{ab}^{(1)},\phi_{ab}^{(2)},\phi_{ab}^{(3)})$ & $K_{\R_a}$ & $\delta_b^{{\rm loop},\mathcal{A}} {\rm f}_{SU(N_a)}$
\\\hline\hline
(0,0,0) &
\begin{tabular}{c} $K_{\Adj_a} = \frac{\sqrt{2 \pi}}{c_a} \, \frac{f(S,U)}{v_i} \sqrt{\frac{V_{aa}^{(j)} V_{aa}^{(k)}}{V_{aa}^{(i)}}}$
\\\hline$K_{\R_a \neq \Adj_a} =f(S,U) \;   \sqrt{\frac{ 2 \pi V_{ab}^{(i)}}{v_jv_k}}$ \\ {\small $(ijk) \simeq (1,2,3)$ cyclic} \end{tabular}
& $\begin{array}{c}- \sum_{i=1}^3 \frac{b_{ab}^{{\cal A},(i)}}{4\pi^2} \ln \eta(iv_i) \\ -\sum_{i=1}^3  \frac{\tilde{b}_{ab}^{{\cal A},(i)}(1-\delta_{\sigma^i_{ab},0}\delta_{\tau^i_{ab},0})}{8\pi^2}
 \times \\ \times \ln \Bigl(e^{-\frac{\pi (\sigma^i_{ab})^2 v_i}{4}}\frac{\vartheta_1 (\frac{\tau^i_{ab} - i \sigma^i_{ab} v_i}{2},i v_i)}{\eta (i v_i)} \Bigr) \end{array}$
\\\hline
\begin{tabular}{c} $(0^{(i)},\phi^{(j)},\phi^{(k)})$ \\ {\footnotesize $\phi^{(j)}=-\phi^{(k)}$} \end{tabular} & $f(S,U) \;   \sqrt{\frac{ 2 \pi V_{ab}^{(i)}}{v_jv_k}}$ 
&$\begin{array}{c} - \frac{b_{ab}^{\cal A}}{4\pi^2} \ln \eta(iv_i)
 - \frac{\tilde{b}_{ab}^{\cal A}(1-\delta_{\sigma^i_{ab},0}\delta_{\tau^i_{ab},0}) }{8\pi^2}\times  \\
\ln \Bigl(e^{-\frac{\pi (\sigma^i_{ab})^2 v_i}{4}}\frac{\vartheta_1 (\frac{\tau^i_{ab} - i \sigma^i_{ab} v_i}{2},i v_i)}{\eta (i v_i)} \Bigr) 
\\ + \sum_{l=j,k} \frac{N_b\,I_{ab}^{\Z_2^{(l)}}}{8 \pi^2 \, c_a} \bigl(\frac{\sgn(\phi^{(l)}_{ab})}{2} -\phi^{(l)}_{ab}) \bigr)
\end{array}$
\\\hline
\begin{tabular}{c} $(\phi^{(1)},\phi^{(2)},\phi^{(3)})$ \\  {\footnotesize $\sum_{i=1}^3 \phi^{(i)}=0$} \end{tabular}&
$f(S,U)  \,  \sqrt{\prod_{i=1}^3 \frac{1}{ v_i} 
\left(\frac{\Gamma(|\phi^{(i)}_{ab}|)}{\Gamma(1-|\phi^{(i)}_{ab}|)}\right)^{-\frac{\sgn(\phi^{(i)}_{ab})}{\sgn(I_{ab})} }}$ 
& $\sum_{i=1}^3 \frac{N_b\, I_{ab}^{\Z_2^{(i)}}}{8 \pi^2 \, c_a}  \left( \frac{\sgn(\phi^{(i)}_{ab}) +\sgn(I_{ab})}{2} -\phi^{(i)}_{ab}\right)$
\\\hline
\end{tabular}
\caption{K\"ahler metrics $K_{{\bf R}_a}$ of open string matter states on D6-branes on $T^6$ and $T^6/\Z_{2N}$ and $T^6/\Z_2 \times \Z_{2M}$ without and
with discrete torsion in dependence of the supersymmetric intersection angles $(\vec{\phi}_{ab})$
containing the universal factor $f(S,U)=(S\, \prod_{i=1}^{h_{21}^{\rm bulk}} U_i)^{-\alpha/4}$ as well as the corresponding annulus contributions 
$\delta_b^{{\rm loop},\mathcal{A}} {\rm f}_{SU(N_a)}$ to the holomorphic gauge kinetic functions as classified in~\cite{Honecker:2011sm}.
The M\"obius strip contributions to the gauge kinetic functions from $aa'$ sectors on $T^6/\Z_{2N}$ are given in 
table~\protect\ref{Tab:Moebius-hol_gauge_kin}.}
\label{Tab:Kaehlermetrics+delta_f_Annulus}
\end{table}
While the K\"ahler metrics and annulus contributions to the holomorphic gauge kinetic functions are fully classified by the vanishing of some
intersection angle $(\phi_{ab}^{(i)})$, and the non-trivial orbifold background dependence 
is absorbed in the beta function coefficients, the M\"obius strip contributions depend also on the relative angles $(\phi_{a,\OR}^{(i)})$ with respect to the 
O6-planes as displayed  in table~\ref{Tab:Moebius-hol_gauge_kin} for $T^6/\Z_{2N}$. 
\begin{table}
\begin{tabular}{|c||c||c||c|}\hline
$(\phi_{aa'}^{(1)},\phi_{aa'}^{(2)},\phi_{aa'}^{(3)})$ &  $\delta_{a'}^{{\rm loop},\mathcal{M}} {\rm f}_{SU(N_a)}$
& $(\phi_{aa'}^{(1)},\phi_{aa'}^{(2)},\phi_{aa'}^{(3)})$ &  $\delta_{a'}^{{\rm loop},\mathcal{M}} {\rm f}_{SU(N_a)}$
\\\hline\hline
$\!\!\!\!\!\!\begin{array}{c} (0,0,0)\; {\rm or} \\ (\phi,0,-\phi)  \\ \pp \OR  \; {\rm on} \; T^2_{(2)}  \end{array}\!\!\!\!\!\!$
& $\!\!\!\!\begin{array}{c} -\frac{b_{aa'}^{\cal M}}{4\pi^2} \ln \eta(i \tilde{v}_2) + \tilde{c}_{\phi} \,\frac{\ln (2)}{2\pi^2} 
\\ - \frac{\tilde{b}_{aa'}^{\cal M} \; (1-\delta_{\sigma^2_{aa'},0}\delta_{\tau^2_{aa'},0})}{8\pi^2} \times \\ 
\ln \Bigl(e^{-\frac{\pi (\sigma^2_{aa'})^2 \tilde{v}_2}{4}}\frac{\vartheta_1 (\frac{\tau^2_{aa'} - i \sigma^2_{aa'} \tilde{v}_2}{2},i \tilde{v}_2)}{\eta (i \tilde{v}_2)} \Bigr) \end{array}\!\!\!\!\!\!$
&  $\!\!\begin{array}{c}i=2 \; {\rm and} \; \perp \OR \; {\rm on} \; T^2_{(2)}\\\hline 
(0^{(i)},\phi^{(j)},\phi^{(k)})_{\phi^{(j)}=-\phi^{(k)}} \\\hline i=1 \; {\rm or} \; 3 \end{array}\!\!$
&$\!\!\begin{array}{c} \frac{\ln (2)}{16\pi^2} \bigl( |\tilde{I}_a^{\OR}| +|\tilde{I}_a^{\OR\Z_2^{(2)}}| \bigr)
\\\hline  -\frac{b_{aa'}^{{\cal M},(i)}}{4\pi^2} \ln \eta(i \tilde{v}_i) 
\\ + \frac{\ln (2)}{16\pi^2} \bigl( |\tilde{I}_a^{\OR}| +|\tilde{I}_a^{\OR\Z_2^{(2)}}| \bigr)
\end{array}\!\!\!\!$
\\\hline
$\!\!\!\!\!\!\begin{array}{c} (0,0,0)\\ \pp \OR\Z_2^{(k), \; k=1\, {\rm or} \, 3} 
\end{array}\!\!\!\!\!\!$
&$ \!\!\begin{array}{c} -\sum_{i=1,3} \frac{b_{aa'}^{{\cal M},(i)}}{4\pi^2} \ln \eta(i \tilde{v}_i)\\
+\frac{1}{8\pi^2} \ln \Bigl(2^4 \,\frac{v_1v_3V_{aa'}^{(1)}V_{aa'}^{(3)} }{(v_2 V_{aa'}^{(2)})^2}\Bigr)
\end{array}\!\!$
& \begin{tabular}{c}  $(\phi^{(1)},\phi^{(2)},\phi^{(3)})$  \\  {\footnotesize $\sum_{i=1}^3 \phi^{(i)}=0$} \end{tabular}
& $\!\!\frac{\ln (2)}{16\pi^2} \bigl( |\tilde{I}_a^{\OR}| +|\tilde{I}_a^{\OR\Z_2^{(2)}}| \bigr)\!\!$ 
\\\hline
\end{tabular}
\caption{Classification of the M\"obius strip contributions $\delta_{a'}^{{\rm loop},\mathcal{M}} {\rm f}_{SU(N_a)}$
to the holomorphic gauge kinetic functions on D6-branes on $T^6/\Z_{2N}$ with 
$\tilde{c}_{\phi=0}=1$  and $\tilde{c}_{\phi\neq 0}=0$  in the upper left entry, see~\cite{Honecker:2011sm} for the full classification on all 
other torus and orbifold backgrounds.}
\label{Tab:Moebius-hol_gauge_kin}
\end{table}

\noindent{\bf Holomorphic gauge kinetic functions for massless U(1) gauge factors.}
Besides the $SU(N_a)$ gauge groups discussed above, $U(1)_a$ factors are ubiquitous in D-brane models and required for model building. 
While the tree-level gauge coupling of a single (anomalous) $U(1)_a$ is simply related to the $SU(N_a) \subset U(N_a)$ part 
by the normalisation factor $2N_a$,
\begin{eqnarray}
{\rm f}^{\rm tree}_{U(1)_a} = 2 N_a \; {\rm f}^{\rm tree}_{SU(N_a)} 
,\qquad
\delta_{\rm total}^{\rm loop} \, {\rm f}_{U(1)_a} = 
 2 N_a \Bigl( 2\, \delta_{a'}^{{\rm loop},{\cal A}}  {\rm f}_{SU(N_a)} +  \delta_{a'}^{{\rm loop},{\cal M}}  {\rm f}_{SU(N_a)}
+ \sum_{b\neq a,a'} \delta_b^{{\rm loop},{\cal A}}   {\rm f}_{SU(N_a)} \Bigr)
,\nonumber
\end{eqnarray}
the one-loop correction differs in the $aa$ and $aa'$ sectors due to their vanishing and doubled $U(1)_a$ charge assignments, respectively~\cite{Honecker:2011sm}. 

The holomorphic gauge kinetic function of  an (anomaly-free) massless linear combination of U(1)s, $Q_X = \sum_i x_i \, Q_{U(1)_i}$, such as the 
SM hyper charge, consists of a superposition of terms from the individual $U(1)_i$ factors plus pairwise mixing terms~\cite{Honecker:2011sm},
\begin{eqnarray}
{\rm f}^{\rm tree}_{U(1)_X} = \sum_i x_i^2 \;  {\rm f}^{\rm tree}_{U(1)_i} 
,\quad
\delta_{\rm total}^{\rm loop} \, {\rm f}_{U(1)_X} = \sum_i x_i^2  \,\delta_{\rm total}^{\rm loop} {\rm f}_{U(1)_i} 
+ 4 \sum_{i<j} x_i x_j N_i \left(\delta_{j^{\prime}}^{{\rm loop},{\cal A}} \, {\rm f}_{SU(N_i)}- \delta_{j}^{{\rm loop},{\cal A}} \, {\rm f}_{SU(N_i)} \right)
.\nonumber
\end{eqnarray}
This formula describes the potential one-loop kinetic mixing of the SM $U(1)$ with some hidden  $Z'$ boson from the open string sector
on D6-branes.

\noindent{\bf The Standard Model example  with hidden $Sp(6)_{h}$ on fractional D6-branes   on $T^6/\mathbb{Z}_6'$}~\cite{Gmeiner:2007zz}{\bf .}
The right-handed quarks $Q_R^{1,2,3}$ of this model are split into two generations $Q_R^{1,2}$ localised at 
intersections of the QCD and `right' stack $ac$ at angle $\pi(0,\frac{1}{2},\frac{-1}{2})$ and with K\"ahler metrics 
$K_{Q_R^{1,2}}=f(S,U) \,\sqrt{\frac{4\pi}{\sqrt{3} \,v_2v_3}}$, whereas the third generation $Q_R^3$ is localised at an 
$a(\theta^2 c)$ intersection at angle $\pi(\frac{-1}{3},\frac{-1}{6},\frac{1}{2})$ and has the K\"ahler metric 
\mbox{$K_{Q_R^{3}}=f(S,U) \,\sqrt{\frac{10}{v_1v_2v_3}}$}.
The three right-handed lepton generations $L_R^{1,2,3}$ are localised at $c(\theta d)$ intersections of the `right' and `leptonic' stack at angle 
$\pi(\frac{-1}{6},\frac{-1}{3},\frac{1}{2})$ and have the K\"ahler metrics $K_{L_R^{1,2,3}}=f(S,U)\,\sqrt{\frac{10}{v_1v_2v_3}}$, cf.~\cite{Honecker:2011sm}.

The tree-level QCD coupling is proportional to the one bulk complex structure $U$ of $T^6/\mathbb{Z}_6'$, and the one-loop correction 
is a sum over contributions from all
D6-branes $a,b,c,d,h$~\cite{Honecker:2011sm},\linebreak
$
\delta_{\rm total}^{\rm loop} {\rm f}_{SU(3)_a} =
 \frac{1}{2\pi^2} \ln \left[ \frac{\eta(i \tilde{v}_1)}{\eta(iv_1) } \frac{1}{\eta(i v_3)^4}
\left(e^{-\pi v_3/4}\frac{\vartheta_1 (\frac{1 - i v_3}{2},i v_3)}{\eta (i v_3)} \right)^{-3/2}
 \left(2^{15/2} \frac{\sqrt{3} \, v_1v_3}{v_2^2} \, r\right)^{1/4}_{r=1/\sqrt{3}}
 \right],$
together with their orbifold and orientifold images, where the orbifold invariant orbit of $a$ is of the special type $\pp \OR\Z_2^{(1)}$ in table~\ref{Tab:Moebius-hol_gauge_kin}.

\section{Conclusions}

By comparison of CFT results with canonical supergravity expressions, we obtained the perturbatively exact holomorphic gauge kinetic 
functions for $SU(N)$ and anomaly-free $U(1)$ gauge groups as well as the tree-level K\"ahler metrics for charged matter on 
D6-branes in $T^6/\Z_{2N}$ orientifold backgrounds, all of which are required for an extension of D-brane model building from 
massless particle spectra to their effective low-energy field theory.
As examples, we explicitly computed the K\"ahler metrics for right-handed quarks and leptons and derived the one-loop holomorphic
gauge kinetic function for the QCD gauge factor of the SM with `hidden' $Sp(6)_h$ on $T^6/\Z_6'$.


\end{document}